\documentclass[12pt,oneside]{amsart}

\usepackage{natbib}
 \bibpunct[, ]{(}{)}{,}{a}{}{,}%
 \def\bibfont{\small}%

\usepackage[margin=1.3in]{geometry}
\usepackage{graphicx}
\usepackage{caption}
\usepackage{subcaption}
\usepackage{latexsym}
\usepackage[T2A]{fontenc}
\usepackage[utf8]{inputenc}
\usepackage[russian,english]{babel}
\usepackage{setspace}
\usepackage{xcolor}
\usepackage{csquotes}
\usepackage{hyperref}
\usepackage{algorithm}
\usepackage[noend]{algpseudocode}

\begin{document}
\thispagestyle{empty}

\title[Bid Leakage in Procurement Auctions]{Stealed-bid Auctions: Detecting Bid Leakage via Semi-Supervised Learning}

\author{Dmitry I. Ivanov}
\address{National Research University Higher School of Economics, 16, Soyuza Pechatnikov st., St.~Petersburg, 190121, Russia} \email{diivanov@hse.ru}

\author{Alexander S. Nesterov}
\email{asnesterov@hse.ru} 

\thanks {Support from the Basic Research Program of the National Research University Higher School of Economics is gratefully acknowledged. 
 We would like to thank Fedor Sandomirskiy, Nikita Kalinin and the entire St.Petersburg Game Theory Lab for useful comments, Sergei Izmalkov and Alexey Drutsa for constructive criticism, and Mikhail Akimov for the open-access script which provided us with initial insights on how to parse the ftp server.\\The paper was previously titled "Identifying bid leakage in procurement auctions: Machine learning approach".}
 

\begin{abstract}
    

Bid leakage is a corrupt scheme in a first-price sealed-bid auction in which the procurer leaks the opponents' bids to a favoured participant. The rational behaviour of such participant is to bid close to the deadline in order to receive all bids, which allows him to ensure his win at the best price possible. While such behaviour does leave detectable traces in the data, the absence of bid leakage labels makes supervised classification impossible. Instead, we reduce the problem of the bid leakage detection to a positive-unlabeled classification. The key idea is to regard the losing participants as fair and the winners as possibly corrupted. This allows us to estimate the prior probability of bid leakage in the sample, as well as the posterior probability of bid leakage for each specific auction.

We extract and analyze the data on 600,000 Russian procurement auctions between 2014 and 2018. We find that around 9\% of the auctions are exposed to bid leakage, which results in an overall 1.5\% price increase. The predicted probability of bid leakage is higher for auctions with a higher reserve price, with too low or too high number of participants, and if the winner has met the auctioneer in earlier auctions.



\end{abstract}

\maketitle

%



\section{Introduction}\label{sec-introduction}

In each country public procurement is an important and complex sector of the economy. In 2017 in Russia, the annual total volume of public procurement was 36.5 trillion rubles, which amounts to around a third of the annual GDP. A majority of contracts in Russian procurement are awarded through auctions, which in theory allocates the contract to the most efficient firm at the lowest possible price. In practice, however, certain tacit manipulations can corrupt the outcome both in terms of efficiency of allocation and the contract price.

In this paper we study `requests for quotations` -- small and frequent first price sealed-bid procurement auctions. These auctions can suffer from \textbf{bid leakage} -- the corruption scheme where the procurer illegally provides his favored participant with the information about the bids of the other participants so that he can respond optimally. Our goal is to estimate how widespread bid leakage is in general and to determine how likely it is that each particular auction has been affected by bid leakage.

We analyze the dataset containing more than 600000 Russian requests for quotations. The dataset covers all the auctions that took place from January 2014 to March 2018 and is extracted from the online database.\footnote{ftp://ftp.zakupki.gov.ru/}

\begin{center}
\begin{figure}[h]
\caption{Example of typical request for quotations with leaked bids}
\center{\includegraphics[scale=0.85]{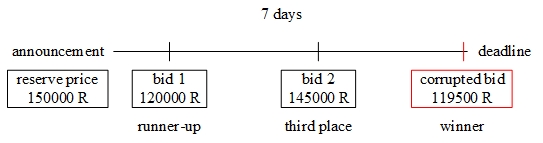}}
\label{fig:image}
\footnotesize
\begin{flushleft}{\textit{Notes}: The auction lasts more than 7 days. The auction is suspicious for bid leakage as the winner bids near the deadline, after every other bid, and only slightly below the runner-up.}
\end{flushleft}
\end{figure}
\end{center}
\normalsize

Our work is inspired by \cite{andreyanov2016corruption} who observed the patterns that are likely to reflect rational behavior of the favoured participant that received leaked bids (see Figure~1): these participants are \\
(1) bidding last, \\
(2) bidding close to the deadline, and \\
(3) winning by a small margin.

The intuition behind these three patterns in auctions with bid leakage is straightforward. First, the only way for the unfair participant to know every other bid and ensure her win is to bid the last, hence pattern (1). Similarly, she delays bidding as much as possible to lower the risks of not being the last, hence pattern (2). As she aims for the highest profit, she slightly undercuts the current best bid, hence pattern (3).


We use these three patterns to determine whether a particular auction has been corrupted by bid leakage. To this end, we develop a two-stage identification strategy.

In the first stage we build a classifier that distinguishes the winners from the runners-up by using features associated with patterns (1), (2) and (3). For a given auction winner, the higher is the predicted probability of winning, the more suspicious the auction is.

In a world without bid leakage, and assuming that these features are not related to the actual bid, such classifier would fail and if it does not then it has to be due to bid leakage. In practice however, the classifier might still be able to predict the winners well even in auctions without bid leakage, which leads to biased estimates. To correct our estimates we construct a synthetic placebo dataset of fair auctions and estimate the sign and the size of the bias.

In the second stage we use the classifier's predictions and performance to estimate the prior probability that a random auction in the dataset is corrupted, and the posterior probability that a specific auction is corrupted -- conditional on the probability of winning that the classifier has assigned to its winner.

We estimate the prior probability of bid leakage as 9\%. We also find that the bid leakage is more likely in auctions with a higher reserve price and when the number of bidders is below 4 or above 7.

The rest of the paper is organized as follows.
In Section \ref{sec-problem-setup} we present the background on requests for quotations and the relevant literature and sketch our identification strategy. In Section \ref{sec-data} we describe the dataset and present the preliminary analysis. In Section \ref{sec-estimation} we present the two stages of our bid leakage estimation: the classifier and the estimates for the prior and the posterior probability of bid leakage. In Section \ref{sec-results} we present the results of estimation and their economic implications. In Section \ref{sec-validation} we validate the method with several indirect tests. Section \ref{sec-conclusions} concludes.


\section{Problem setup and identification strategy}\label{sec-problem-setup}

\subsection{Requests for quotations and background on bid leakage detection}\label{background}

In Russia requests for quotations are used for distributing small contracts such as roof repair for a factory or products delivery to a school kitchen. Before each auction starts, the procurer makes an announcement with the relevant information about the contract and the auction. The announcement includes reserve price -- the maximal price the contract can be assigned for, the reserve price is bounded by 500000 rubles (approximately \$8000). The auction lasts at least one week. During this period potential participants can submit their bids. Each participant can submit only one bid, the bids are sealed. After the auction ends, all bids are revealed and the smallest bid wins, the final price equals the winning bid (first-price auction).


The literature on manipulations in auctions is prolific but mostly concerned with collusion schemes such as bid rigging \citep{porter1993detection,imhof2018screening} and bid rotation \citep{aoyagi2003bid}; a review of the literature on collusion detection is available in \citep{harrington2005detecting}.

Crucial to our research question is the timing of bids, and the Russian procurement data is unique to contain this information.  Previously, timing of bids has been studied in repeated Internet auctions such as eBay, where each bidder has a set of moments in time where he can submit a bid \citep{song2004nonparametric}. But, to the best of our knowledge, timing of bids has not been used before to detect corruption (except for \cite{andreyanov2016corruption} that we discuss below).

Other papers studying bid leakage or other forms of corruption using Russian data do so on a local scale of a specific market, during a specific period of time or employing the state-ownership of bidders \citep{yakovlev2015incentives, mironov2016corruption, balsevich2014indicators,tkachenko2017sweet}. The empirical literature on auctions is rich (see, e.g., the seminal works by \cite{athey2007nonparametric} and \cite{krasnokutskaya2011bid}), yet there are only few studies that use supervised machine learning (classification) in auction corruption detection. Typically they utilize small datasets of few hundreds of labeled auctions \citep{ferwerda2017corruption,huber2018machine}.

\subsection{Identification strategy and placebo}\label{sec-identification}

The only closely related papers to ours are \citep{andreyanov2016corruption} and its recent version \citep{korovkin2018detecting}: we study the same object using the same data. However, our identification and estimation strategy is different in three crucial ways: we use weaker assumptions, more advanced methods and larger set of characteristics (i.e., features), which additionally enables us to determine posterior probability of bid leakage for each specific auction.

The identification in \citep{andreyanov2016corruption,korovkin2018detecting} relies on a crucial assumption that, in the auctions without bid leakage, bids and timing of the bids are independent. If \textit{independence} holds, then the higher likelihood of the last bids to win compared to earlier bids is attributed exclusively to bid leakage.

The \textit{independence} assumption might fail for a number of reasons. The longer time it takes a risk-averse bidder to study the case, the lower will his bid be. For example, in our data we observe that  1,2\% of participants  behave like `snipers`: they bid during the first day of the auction and bid slightly below the reserve price (up to 5\%).

Another reason comes from the honest bidders' attempts to resist bid leakage. On the one hand, the later the bid is submitted, the lower is the chance that it is going to be leaked and undercut by a corrupted bidder. On the other hand, submitting closer to the deadline requires attention and possibly costly logistics.\footnote{Anecdotal evidence suggests that some bidders did not rely on post or courier services and delivered their bids personally to make sure to submit just before the deadline.} As a result, a bidder with a higher valuation (i.e., lower execution costs) submits a lower bid and, simultaneously, is ready to delay the submission more relative to the bidders with the lower valuations. Because bidding and timing are confounded through valuation, the \textit{independence} assumption does not hold. A later bid has a higher chance of winning not only due to presence of bid leakage, but due to mere expectation of bid leakage. We present this argument formally in the  Subsection \ref{sec-gt-model}.

We do not rely on \textit{independence} assumption. Instead, we correct our estimates using a synthetic placebo dataset of fair auctions. We remove the first-ranked bidders (the true winners) from all the auctions and recalculate the features accordingly. This way, we obtain a new dataset where the second places are treated as the winners and the third places -- as the runners-up (Figure 2). We estimate the bias in these synthetic auctions and assume that this bias is equal to the bias in fair auctions in the original dataset. We verify this assumption and compare it to \textit{independence} in Section \ref{sec-verify-assumption}.

\begin{figure}[h]
\caption{Placebo auction example, transformed from the auction at Figure 1}
\center{\includegraphics[scale=0.85]{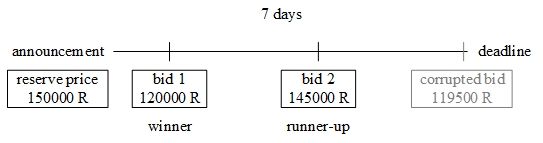}}
\footnotesize
\begin{flushleft}{\textit{Notes}: The placebo auction is generated by removing the winning bid from the auction. In this hypothetical auction, bid 1 belongs to the winner, bid 2 -- to the runner-up. The auction is no longer suspicious for bid leakage.}
\normalsize
\end{flushleft}
\end{figure}

Second, in contrast to reduced-form and structural statistical models we use machine learning techniques. This allows us to consider all the evidence on bid leakage patterns at once and without imposing restrictions on their interconnection. When we only include subset of the relevant features as it is done in \citep{korovkin2018detecting}, the method identifies significantly less bid leakage.

Finally, we descend from the population level and develop a method able to assign the posterior probability of the bid leakage presence to every auction in the dataset. Our method provides more precise and specific estimates of bid leakage and can be used for automatic ex-post bid leakage detection, which can be useful for regulation and auditing authorities.

\subsection {Game-theoretic model of bid leakage}\label{sec-gt-model}

In this section, using a simple game-theoretic model, we show why \textit{independence} between bids and timing might fail even without bid leakage. 

First consider the world without bid leakage. An auctioneer is selling a procurement contract with reserve price normalized to 1, and the lowest possible cost of executing the contract is normalized to zero. Each bidder is risk-neutral and is drawing his execution costs $e$ from a uniform distribution on $[0,1]$, or, equivalently, each bidder has an iid valuation $v=1-e$ drawn from a uniform distribution on $[0,1]$; the cumulative density function (cdf) is $F(v)=v$.

Let the expected number of bidders participating in the auction be $n$.\footnote{Since bidders are risk-neutral and valuations are i.i.d. the uncertainty regarding the exact number of bidders does not play a role as shown by \cite{matthews1987comparing,mcafee1987auctions}.} For each bidder $i$ with a valuation $v$ his equilibrium bid is the expected bid of his runner-up conditional on $i$ being the winner, $$b^*(v)=1-\frac{\int_{0}^{v}xdF^{n-1}(x)}{\int_{0}^{v} dF^{n-1}(x)}=1-v\frac{n-1}{n}.$$
Since the bid is monotonic in the valuation, in order to win the bidder needs to have the highest valuation. The probability of winning is thus $F^{n-1}=v^{n-1}$, and the expected profit is $$E\pi(v,b^*(v))=\frac{v^n}{n}.$$

Now we add the time dimension to the problem. Each bidder chooses the submission time $t\in [0,1]$. We assume that delaying submission is costly, submitting at time $t$ costs the bidder $c(t)$, where $c$ is increasing and convex, $c'>0,c''>0$, and extremely high close to deadline, $c(1)=\infty$. These costs represent the stress and attention costs of not missing the deadline and also the costs of more precise bid delivery.\footnote{In fact, the Federal Antimonopoly Service of the Russian Federation reports a number of official complaints from the bidders that their bids were illegally rejected due to (claimed) late submission, see the database at \url{br.fas.gov.ru}.} In the world without bid leakage the timing of the bid is irrelevant for winning and each bidder submits at time $t=0$.

Now let the bid leakage be possible. We assume that each bidder has a common prior belief regarding the possibility of bid leakage. Conditional on that the auction is corrupted, the probability that a specific bid is leaked and undercut decreases in time of submission: if you submit later, then the chances that your bid is leaked are lower. We assume that for each bidder the perceived probability that his bid submitted at time $t$ is leaked and undercut is exogenously given by some function $\beta(t)$ decreasing in time, $\beta'(t)<0$, down to zero at the time of deadline $t=1$, $\beta(1)=0$. \footnote{The implicit simplifying assumption here is that the favored bidder has lower costs than each bid that is leaked is undercut. This makes the honest bidders counter bid leakage only by submitting bids later but not by decreasing their bids.}

Thus the expected profit of the bidder with valuation $v$ and bid $b^*(v)$ is as follows: $$E\pi(v,b^*(v),t)=\frac{v^n}{n}(1-\beta(t)) - c(t).$$ The optimal submission time $t^*(v)$ is given by the first order condition: $$c'(t^*(v))=-\frac{v^n}{n}\beta'(t^*(v)).$$

Observe that both $b^*(v)$ is decreasing and $t^*(v)$ is increasing in valuation $v$. Thus the optimal bid and the submission time of the bid are confounded by the valuation. The higher the valuation is, the more is the bidder ready to pay to get the contract: both in terms of submitting a lower bid and in terms of costly delay of the submission.

Observe also that the correlation between the timing and the bid holds for each valuation, and thus will be true not only for the winners but also for the runners-up. We uncover this correlation for runners-up using a placebo dataset in Section \ref{sec-second-stage} and use it to correct our biased estimates for the winners.

\subsection{Positive-Unlabeled Classification}\label{sec-pu-description}

Our bid leakage estimation strategy is based on Positive-Unlabeled (PU) classification. Generally, PU classification is applied instead of supervised classification in the cases when the training dataset is not fully labeled. Specifically, only a labeled sample from the positive class is assumed to be available, while the rest of the positive data and all the negative data are mixed in the Unlabeled sample. This problem formulation can be applied to our setup if we regard runners-up as fair (positive) and the winners as possibly corrupted (unlabeled). Other applications of PU learning include detection of fake texts \citep{ren2014reviews}, classification of time-series \citep{nguyen2011ts}, and identification of disease genes \citep{yang2012desease}.

In the last two decades, numerous methods have been proposed to tackle PU classification \citep{liu2002spy,lee2003weightedLR,en,nnRE}. The technique that achieves the state-of-the-art performance is DEDPUL \citep{ivanov2019}. This method is able to both estimate the proportion of positives in the unlabeled sample and classify it. The details about DEDPUL and its modification that incorporates our assumptions about the domain are described in Section \ref{sec-estimation}.

\section{Auction Data}\label{sec-data}

\begin{table}
\centering
\small
\caption{dataset characteristics}
\label{table-data}

\begin{tabular}{|c|c|c|c|}
\hline
\textbf{Characteristics}                                                               & \textbf{Mean} & \textbf{Median} & \textbf{Std. Dev.} \\ \hline
Number of participants                                                                 & 4.2             & 4               & 1.9                \\ \hline
Reserve price, rubles                                                                  & 199000        & 155000          & 151000             \\ \hline
Winner's bid, rubles                                                                   & 133000        & 95000           & 115000             \\ \hline
Runner-up's bid, rubles                                                                & 148000        & 110000          & 133000             \\ \hline
Price fall                                                                             & 0.339          & 0.305            & 0.205               \\ \hline
\begin{tabular}[c]{@{}c@{}}Time from bid\\ to deadline, hours\end{tabular}             & 39.1            & 19.5              & 54.4                 \\ \hline
\begin{tabular}[c]{@{}c@{}}Time from winner's bid\\ to deadline, hours\end{tabular}    & 35.1            & 18.2              & 51.8                 \\ \hline
\begin{tabular}[c]{@{}c@{}}Time from runner-up's bid\\ to deadline, hours\end{tabular} & 37.5            & 19.3              & 52.5                 \\ \hline
Duration, hours                                                                        & 202           & 171             & 74                 \\ \hline
\end{tabular}

\smallskip
\footnotesize
\begin{flushleft}
\textit{Notes}: The dataset is preprocessed as described in Section \ref{sec-data-preprocessing}. We define Price fall as $\frac{r - b_1}{r}$, where r is reserve price, $b_1$ is winner's bid.\
\end{flushleft}
\normalsize
\end{table}

\begin{table}[t]
\centering
\small
\caption{Auctions with different number of bidders}
\label{table-participants}
\begin{tabular}{|c|c|c|c|c|c|c|c|c|c|}
\hline
\begin{tabular}[c]{@{}c@{}}number of \\ participants\end{tabular}            & 1      & 2      & 3      & 4     & 5     & 6     & 7     & 8     & 9       \\ \hline
\begin{tabular}[c]{@{}c@{}}number of\\ auctions\end{tabular}                 & 610451 & 359588 & 137945 & 64297 & 33404 & 18362 & 10405 & 6122  & 3898  \\ \hline
\begin{tabular}[c]{@{}c@{}}share of auctions\\ won at reserve price\end{tabular} & 0.320  & 0.055  & 0.010  & 0.008 & 0.008 & 0.008 & 0.004 & 0.004 & 0.006 \\ \hline
p(win $\mid$ last)                                              & 1      & 0.487  & 0.391  & 0.305 & 0.246 & 0.210 & 0.185 & 0.173 & 0.154 \\ \hline
p(win)                                                                       & 1      & 0.5    & 0.333  & 0.25  & 0.2   & 0.166 & 0.143 & 0.125 & 0.111   \\ \hline
\end{tabular}
\end{table}

\subsection{Preprocessing}\label{sec-data-preprocessing}


We extract data\footnote{The procurement auctions' data is stored at \url{ftp://zakupki.gov.ru}} on 1271477 requests for quotations that took place between January 2014 and March 2018. The data was preprocessed in the following way:

\begin{itemize}
\item The auctions with missing data or with obvious coding mistakes were dropped. These obvious mistakes include: reserve price being negative or higher than the upper bound 500000 rubles; the starting date being after the ending date; the starting date, the ending date or the bidding date being in the future; the bid being negative or higher than the reserve price. Consequently, 98.5\% of the dataset is left.
\item Our identification method can only be applied to the auctions with 3 participants or more. After dropping the auctions with 1 and 2 participants, 22.2\% of the dataset is left.
\item In the cases of tied bids, the earliest bid wins. Such tie-braking procedure may provide unintended ques for the winner's classifier, e.g. an early bid with zero margin is likely to win. To exclude ties from the dataset, all tie-losing bids are dropped. As a result, the number of participants in some auctions decreases below 3, and such auctions become unavailable for our analysis. After this last modification, 21.3\% of the dataset is left.
\end{itemize}

After the preprocessing, 271209 auctions remain in the main sample. The characteristics of this sample are summarized in Table \ref{table-data}. To cover a bigger portion of the initial dataset, we extend our bid leakage estimation strategy to the auctions with 2 participants in Section \ref{sec-results-2participants}. In total, we analyze around 48.3\% of the dataset, mainly excluding the auctions with 1 participant.

\subsection{Exploratory Analysis}\label{data-analysis}

In this section we present the figures illustrating the suspicious patterns in the data.

We begin with the first two patterns anticipated in the presence of bid leakage: the winners bid close to the deadline and they are more likely to be the last to bid. 

\begin{figure}
    \centering
    \begin{subfigure}[t]{0.65\textwidth}
        \centering
        \vspace{1pt}
        \includegraphics[width=\linewidth]{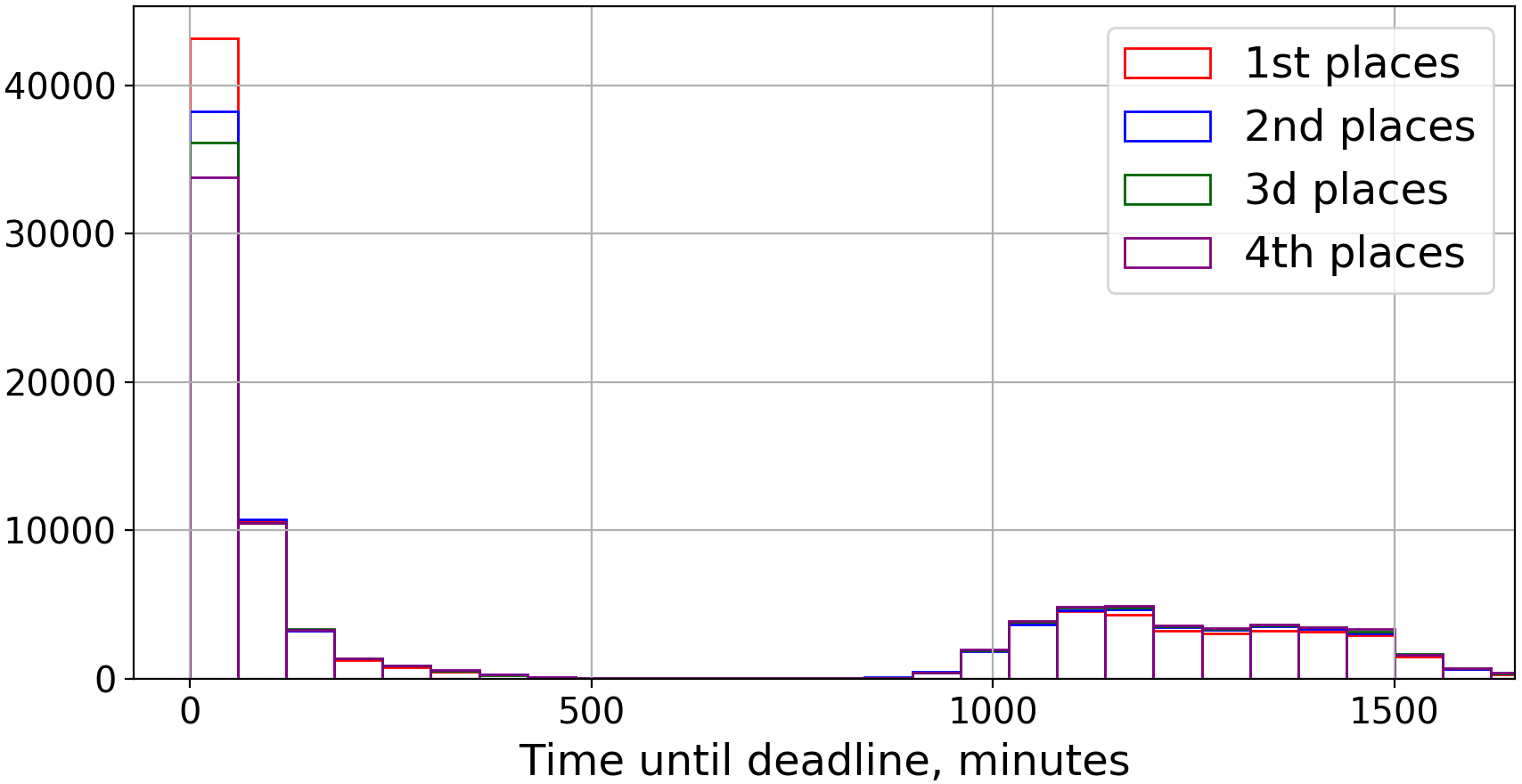} 
        \caption{Timing of bids} \label{fig:timing}
    \end{subfigure}
    \hfill
    \begin{subfigure}[t]{0.3\textwidth}
        \centering
        \vspace{15pt}
        \includegraphics[width=\linewidth]{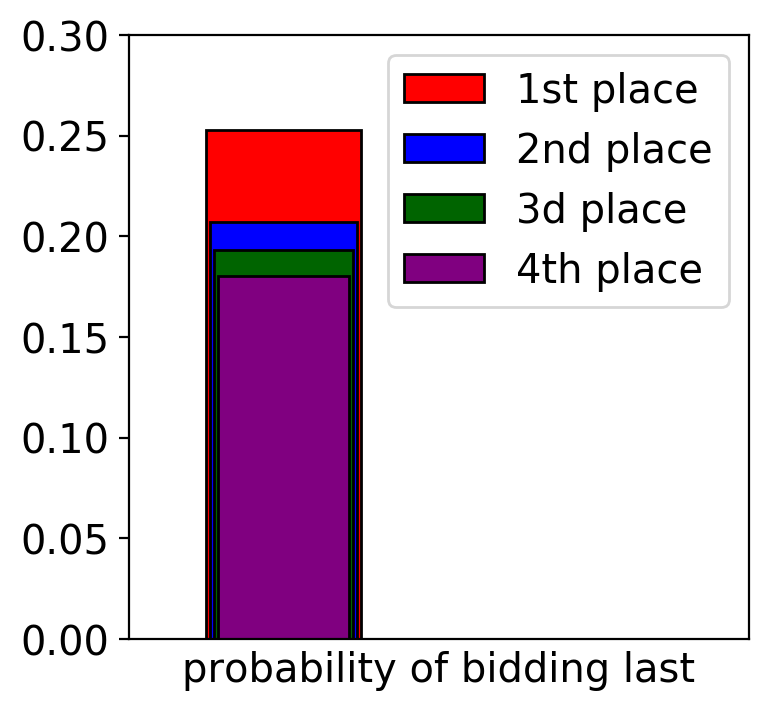} 
        \vspace{-6pt}
        \caption{Order of bids} \label{fig:order}
    \end{subfigure}
    \caption{Timing and order of the bids in auctions with at least 4 bidders}
\footnotesize
\begin{flushleft}{\textit{Notes}: In (\subref{fig:timing}) you can see the distribution of bidding times for the winners, 2nd places and others; time grid is 1 hour. In (\subref{fig:order}) you can see the probability to bid the last for the winners, 2nd places and others.  }
\normalsize
\end{flushleft}
       
\end{figure}

For the auctions with at least 4 bidders we plot the timing of the bids for the winners, the 2nd, the 3rd and the 4th places (Figure \ref{fig:timing}). All bidders are more likely to bid in the last hour before the deadline than in any other hour, but the winners are significantly more likely to do so than others.\footnote{The order of the bidders reflects the intuition from the model in section \ref{sec-gt-model}: as the bids and the timing are confounded by the valuation, the winners are more likely to bid closer to the deadline than the 2nd places, which are more likely than the 3rd and so on. Also, the difference between 2nd and 3rd places is the same as between 3rd and 4th, which illustrates the parity assumption in section \ref{sec-estimation}.}

The winners are also about 5\% more likely to be the last to bid (Figure \ref{fig:order}). This 5\% gap is almost the same for auctions with different number of bidders (Table \ref{table-participants}).\footnote{The only exception are the auctions with 2 bidders, but these are more suspicious for collusion: e.g. in 5\% of these auctions both bids equal the reserve price, and the earlier bid wins due to the tie-breaking rule.}

\begin{figure}\label{bid-margins}
    \centering
    \begin{subfigure}[t]{0.49\textwidth}
        \centering
        \includegraphics[width=\linewidth]{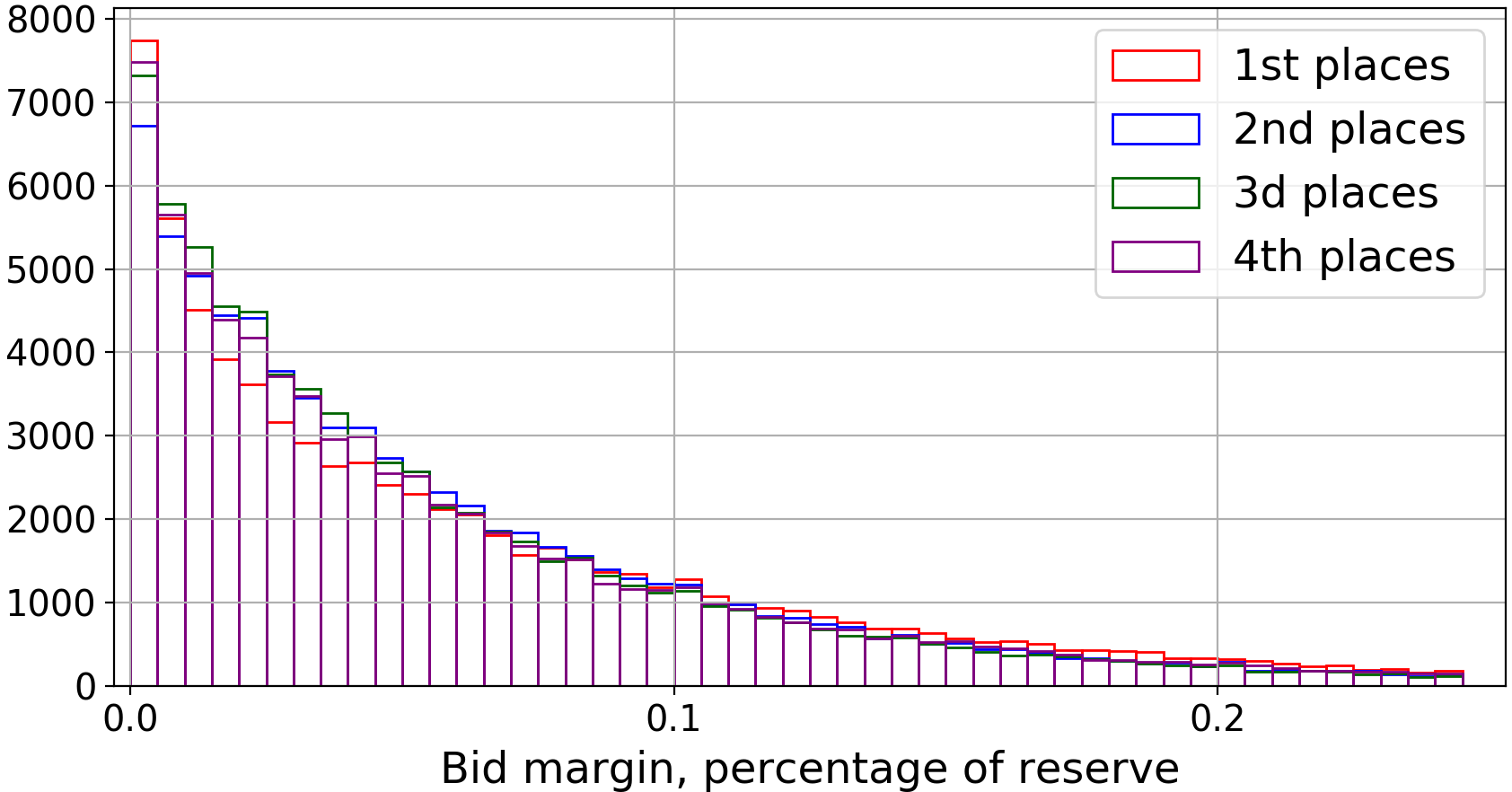} 
        \caption{All auctions} 
        \label{fig:margin-all}
    \end{subfigure}
    \hfill
    \begin{subfigure}[t]{0.49\textwidth}
        \centering
        \includegraphics[width=\linewidth]{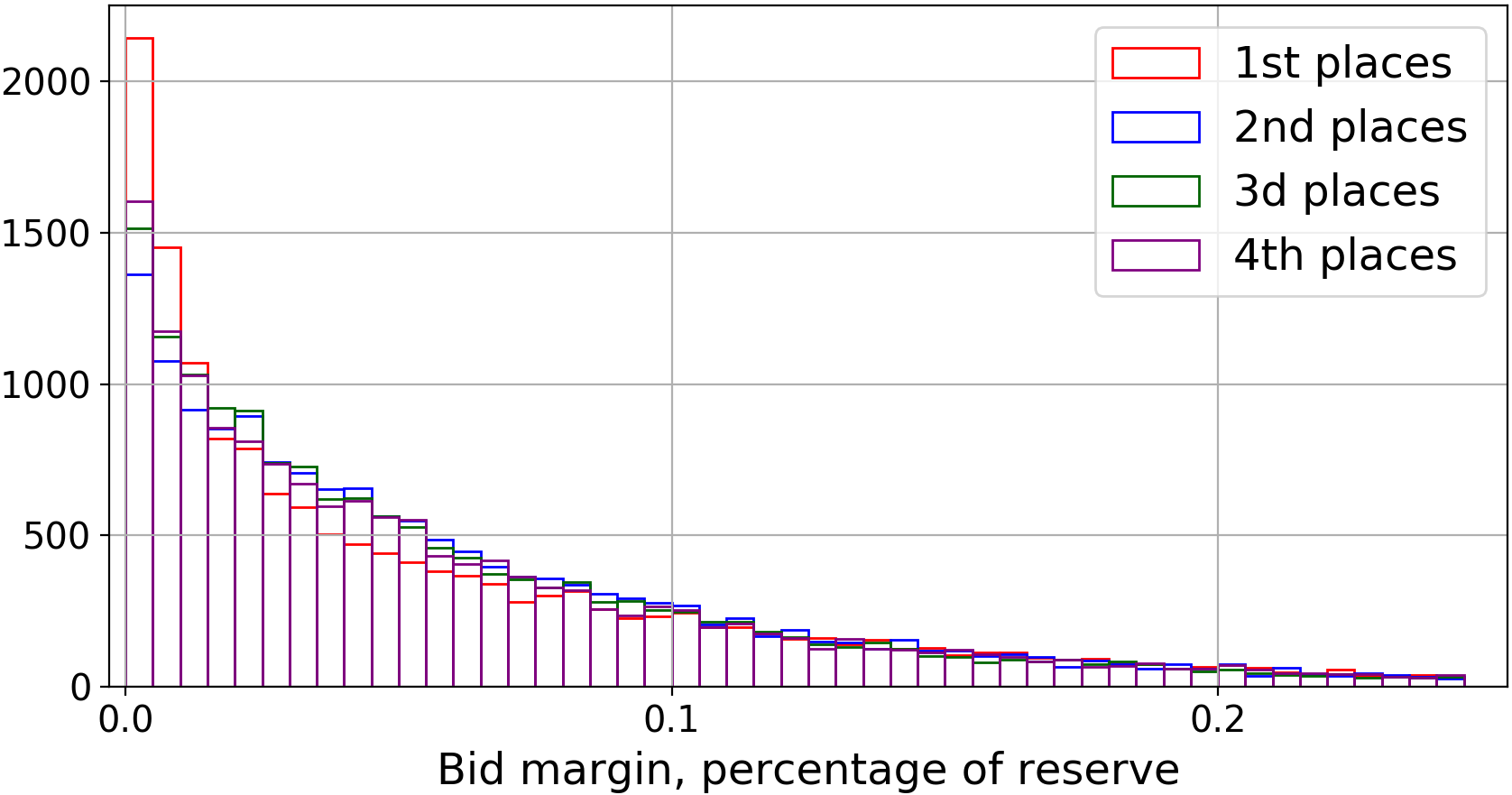} 
        \caption{Auctions with winner bidding last} \label{fig:margin-last-wins}
    \end{subfigure}
    \caption{Margins between consecutive bids in auctions with at least 5 bidders}
\end{figure}

Next we consider the 3rd pattern: in auctions with bid leakage the winners are more likely to undercut --- to win by a small margin. Again, for the auction with at least 5 bidders we plot the bid margins for the first 4 places as a percentage of the reserve price (Figures \ref{fig:margin-all}, \ref{fig:margin-last-wins}).

In the full sample (Figure \ref{fig:margin-all}) we observe 3 stylized patterns for the winners relative to the others:  winners are more likely to have small bid margin (below 0.5\% of the reserve price), less likely to have medium bid margin (below 10\% of the reserve price) and more likely to have large bid margin (above10\% of the reserve price).

When we restrict the sample to the auctions where the winner is bidding last (Figure \ref{fig:margin-last-wins}), these patterns become more pronounced. Here the winners are much more likely to have small bid margin (below 1.5\% of the reserve price), less likely to have medium bid margins and are as likely as others to have larger bid margin. 

The presence of bid leakage can explain these patterns. If the winners bid last, and thus might know all the leaked bids, they are more likely to have very small bid margin; if the winners do not bid last, and thus do not know at least one bid, they are more likely to have a large bid margin.

\section {Bid Leakage Estimation}\label{sec-estimation}

\begin{algorithm}
\caption{Bid leakage estimation}\label{algorithm-strategy}
\small
\begin{algorithmic}[1]

  \State{\textbf{Input:} Samples of winners $X_1$, runners-up $X_2$, placebo winners $X_{1_{p}}$, placebo runners-up $X_{2_{p}}$}
  
  \State{\textbf{Output:} Proportion of corrupted auctions $\alpha$, probability that each winner is corrupted $\{p_{bl}(x): x \in X_1\}$}
  
  \State{Classifier.train($X_1$, $X_2$) $\backslash \backslash$ \emph{a classifier that distinguishes winners from runners-up}}
  
  \State{$Y_1$, $Y_2$ = Classifier.predict($X_1$), Classifier.predict($X_2$)}
  
  \State{$Y_{1_p}$, $Y_{2_p}$ = Classifier.predict($X_{1_p}$), Classifier.predict($X_{2_p}$) $\backslash \backslash$ \emph{predicted probabilities of winning; train and predict are applied to different folds several times in a cross-validation manner}}
  
  \State{$f_1$, $f_2$, $f_{1_p}$, $f_{2_p}$ = kde($Y_1$), kde($Y_2$), kde($Y_{1_p}$), kde($Y_{2_p}$) $\backslash \backslash$ \emph{estimate probability density functions of predictions using kernel density estimation}}
  
  \State{$\Delta_{p}(y) = f_{1_p} - f_{2_p}$ $\backslash \backslash$ \emph{difference of densities in placebo}}
  
  \State{$\Delta(y) = \Delta_p(y)$ $\backslash \backslash$ \emph{apply parity assumption}}
  
  \State{$\alpha = 0$ $\backslash \backslash$ \emph{initialize the priors}}
  
  \Repeat $\backslash \backslash$ \emph{EM algorithm identifies priors and posteriors}
  
    \State{$\alpha_{prev} = \alpha$}
    
    \State{$p_{bl}(y) = \{\max\left[1 - \frac{(1 - \alpha) (f_{2}(y) + \Delta(y))}{f_1(y)}, 0\right]: y \in Y_1\}$ $\backslash \backslash$ \emph{E-step}}
    
    \State{$\alpha = \frac{1}{\left| Y_1 \right|}\sum_{y \in Y_1}\left(p_{bl}(y)\right)$ $\backslash \backslash$ \emph{M-step}}
    
  \Until \emph{$|\alpha_{prev} - \alpha| < tolerance$} $\backslash \backslash$ \emph{we use $tolerance = 1e^{-5}$}
  
  \State{\textbf{Return} $\alpha$, $p_{bl}(y)$ $\backslash \backslash$ \emph{$p_{bl}(y) = p_{bl}(x)$, i.e. classifier preserves the posteriors, as proven in \citep{ivanov2019}}}
  
\end{algorithmic}
\end{algorithm}

Here we describe our two-stage bid leakage estimation strategy. The strategy is summarized in Algorithm \ref{algorithm-strategy}. Mainly, we reduce the problem to Positive-Unlabeled Classification by considering the runners-up (second places) as fair (Positive) participants and the winners as a mixture of fair (Positive) and corrupted (Negative) participants.

We follow the state-of-the-art DEDPUL procedure proposed in \cite{ivanov2019} for general purposes. At the first stage of the procedure a supervised binary classifier is trained to distinguish the winners from the runners-up. Predictions of this classifier are obtained for all the winners in the dataset. Under \textit{independence} or \textit{parity} assumptions, these predictions can be regarded as biased probabilities of bid leakage, as an anomaly score. At the second stage the predictions are transformed into bid leakage probabilities by estimating ratio of probability density functions (pdf) of the predictions for the runners-up and for the winners.

There is a crucial distinction of our strategy compared to the original DEDPUL procedure. The latter would assume that bid leakage is the only reason why the winners and the runners-up differ for the classifier. However, as we have already discussed in Sections \ref{sec-identification} and \ref{sec-gt-model}, this might not be the case, and the difference may exist even in the fair auctions. To account for this we introduce into the analysis the synthetic placebo auctions defined in Section \ref{sec-identification}, which are assumed to be fair, and modify the procedure correspondingly. These modifications are discussed in details in Section \ref{sec-second-stage}.

\subsection{First Stage: Winner vs Runner-up Classifier}\label{sec-classifier}

In the first stage we train the classifier to distinguish the bids of the winners from the bids of the runners-up.

The features that the classifier is trained on (Table \ref{table-features}) are specifically designed to reflect possible bid leakage patterns, while uncovering only little information about fair auctions. Specifically, the features \textit{bid last?} and \textit{bid timing} reflect intention of a corrupted participant to gather information about all the other bids. Small values of \textit{bid margin} reflect undercutting. The feature \textit{met before?} reflects the possibility of repeated procurer-participant cooperation.

\begin{table}
\centering
\small
\caption{Features description}
\label{table-features}
\begin{tabular}{|c|c|c|c|}
\hline
\textbf{Name}                                                    & \textbf{Type} & \textbf{Range} & \textbf{Description}                                                                                           \\ \hline
bid last?                                                        & Binary        & \{0,1\}        & \begin{tabular}[c]{@{}c@{}}Did participant bid\\ after other participants?\end{tabular}                        \\ \hline
met before?                                                      & Binary        & \{0,1\}        & \begin{tabular}[c]{@{}c@{}}Was participant in auction\\ with this procurer before?\end{tabular}                \\ \hline
bid timing                                                       & Continuous    & {[}0, 1440{]}  & \begin{tabular}[c]{@{}c@{}}Minutes from the moment\\ bid is made to deadline\end{tabular}                      \\ \hline
bid margin                                                     & Continuous    & {[}0, 0.05{]}  & \begin{tabular}[c]{@{}c@{}}Difference with bid of succeeding place,\\ normalized by reserve price\end{tabular} \\ \hline
\begin{tabular}[c]{@{}c@{}}number of\\ participants\end{tabular} & Integer       & {[}2, 86{]}    & Number of participants in auction                                                                              \\ \hline
\end{tabular}

\smallskip
\footnotesize
\begin{flushleft}
\textit{Notes}: \textit{bid margin} is clipped from above at 0.05: values bigger than 0.05 are set to this threshold. Likewise, \textit{bid timing} is clipped at 1440 minutes (1 day). This is supposed to cover information about fair auctions. Auctions with 1 participant are excluded from the analysis. 
\end{flushleft}
\normalsize
\end{table}

Note that the information about whether some two participants are from the same auction is purposefully lost. The classifier does not choose the winner between the two participants in each auction. Instead, it determines the chances that each set of features in the dataset belongs to a winner as opposed to a runner-up. As a result, the classifier compresses original multi-dimensional data to one dimension of predicted probabilities. This becomes crucial in the second stage when pdfs of these predictions are explicitly estimated.


As the main classifier we train a feed-forward neural network \citep{pytorch} with 2 layers of 256 neurons via Adam optimizer \citep{adam}. The probability of winning is predicted for all winners and runners-up in the dataset via 5-fold cross-validation. At the second stage we establish the connection between these predictions and the probability of bid leakage.

Additionally, we apply gradient boosting of decision trees \citep{xgboost} as a classifier to check the robustness of the estimates. We train an ensemble of 60 trees with the depth of each tree limited to 5 levels.

\subsection{Second Stage: Transforming Classifier's Predictions into Bid Leakage Probability}\label{sec-second-stage}

We show how to use the winner's classifier to estimate both the prior and the posterior probabilities of bid leakage, neither of which are assumed to be known in advance for any of the auctions. First, we introduce the notations and formally define the problem.

At the first stage, the classifier estimates the probability that a participant wins based on the corresponding vector of features $x$ (Table \ref{table-features}). Denote this probability of winning as $y(x)$. Denote pdfs of distributions of $y(x)$ for winners and runners-up as $f_1(y)$ and $f_2(y)$ respectively. For simplicity, these can be also regarded as pdfs of distributions of initial features $x$.

As was previously discussed, we consider the runners-up to be fair participants (as we aim to detect only successful bid leakage), while the winners may contain both fair and corrupted participants. Moreover, the distributions of $y(x)$ for the winners and the runners-up of the fair auctions might also differ. This may generally be expressed in the following mixture model:

\begin{equation}
    f_2(y) = f_{\overline{bl}_2}(y)
\end{equation}

\begin{equation}
    f_1(y) = \alpha f_{{bl}_1}(y) + (1 - \alpha) f_{\overline{bl}_1}(y) = \alpha f_{{bl}_1}(y) + (1 - \alpha) (f_2(y) + \Delta_{12}(y))
\end{equation}

\begin{equation}
    \Delta_{12}(y) = f_{\overline{bl}_1}(y) - f_{\overline{bl}_2}(y) = f_{\overline{bl}_1}(y) - f_2(y)
\end{equation}

\noindent where $\alpha$ denotes the prior probability of bid leakage; subscript $bl$ denotes bid leakage, i.e. unfair auctions; subscripts $1$ and $2$ denote winners and runners-up; $f_1(y)$ and $f_2(y)$ denote pdfs of classifier's predictions $y(x)$ for winners and runners-up; $f_{bl_1}(y)$, $f_{\overline{bl}_1}(y)$, and $f_{\overline{bl}_2}(y)$ denote pdfs for corrupted winners, fair winners, and fair runners-up respectively; $\Delta_{12}(y)$ denotes difference between pdfs of winners and runners-up in fair auctions.

Equation (1) means that all runners-up are fair. Equation (2) means that winners are a mixture of corrupted and fair participants with mixing proportion $\alpha$. Assuming \textit{independence} would yield $f_{\overline{bl}_1}(y) = f_{\overline{bl}_2}(y)$. Since we find evidence that it does not hold, we introduce $\Delta_{12}(y)$ in the right-hand part of equation (2) instead. It is defined in equation (3) and accounts for the difference between fair winners and runners-up. Introduction of $\Delta_{12}(y)$ is exactly what distinguishes our case from the standard PU Classification problem setup.\footnote{A common assumption in PU literature is that the distributions of labeled and unlabeled positives coincide. It was first introduced as Selected Completely At Random (SCAR) by \cite{en} and is equivalent to our \textit{independence} assumption. An emerging alternative to SCAR is Selected At Random (SAR) \citep{bekker2018beyond}, which allows the labeling probability to be a propensity function $e(x)$, as opposed to being constant. While we also go beyond SCAR, SAR does not fit our setup either. Specifically, the reason SCAR does not hold in our dataset is not that the labeling probability is instance-dependent, but that the mixture distribution $f_1(y)$ (winners) differs from the positive distribution $f_2(y)$ (runners-up) \textit{even} in the absence of corruption ($\alpha = 0$ and hence $\forall x: e(x) = 1$). For this reason, we assume \textit{parity} instead.}

Our goal is to estimate the prior probability $\alpha$ that a random winner is corrupted and the posterior probability $p_{bl}(y)$ that a specific winner is corrupted. The latter may be expressed using the Bayes rule and then transformed using equation (2):

\begin{equation}
    p_{bl}(y) = \frac{\alpha f_{bl_1}(y)}{f_1(y)} = 1 - \frac{(1 - \alpha) (f_{2}(y) + \Delta_{12}(y))}{f_1(y)}
\end{equation}

In equation (4), `ies $f_1(y)$ and $f_2(y)$ can be estimated from the data, e.g. via kernel density estimation. Notice that this procedure does not suffer from the curse of dimensionality, since the distributions are one-dimensional. The difference $\Delta_{12}(y)$ is also unknown, and this issue will be addressed later; for now, consider it exogenous. Then, the only unknowns in (4) are the priors and the posteriors. As proposed in DEDPUL \citep{ivanov2019}, both may simultaneously be estimated by applying Expectation-Maximization algorithm to (4). The procedure is to initialize $\alpha = 0$, estimate posteriors with (4), update the priors as the average posteriors $\alpha = \frac{1}{n} \sum p_{bl}(y)$, and repeat until convergence, i.e. until the priors and the average posteriors become sufficiently close.

We now address the issue of $\Delta_{12}(y)$ estimation. The key step is to construct the synthetic dataset of implicitly fair placebo auctions. As was previously discussed, placebo auctions are generated from the real auctions by dropping the winners and keeping the other participants which we know to be fair. In each hypothetical auction, the second-ranked bidder is assumed to be the winner, and the third-ranked bidder is assumed to be the runner-up.

By applying the classifier that is trained on the real auctions to the placebo dataset, we may obtain its predictions to later estimate pdfs $f_{1_{(2,3)}}$ and $f_{2_{(2,3)}}$ for the winners and the runners-up of placebo auctions respectively (subscript (2,3) implies that the original winners are dropped), where:

\begin{equation}
    \Delta_{23}(y) = f_{1_{(2,3)}}(y) - f_{2_{(2,3)}}(y)
\end{equation}

Thus, we may estimate $\Delta_{23}(y)$ by using placebo auctions. This becomes crucial as we make a major assumption regarding equality of $\Delta_{12}(y)$ and $\Delta_{23}(y)$:

\textbf{PARITY:} $\Delta_{12}(y)$ = $\Delta_{23}(y)$. The difference between probability density functions of distributions of winner's classifier' predictions for winners and runners-up in real fair auctions is equal to this difference in placebo auctions.

Using the \textit{parity} assumption we may estimate $\Delta_{12}(y)$ and thus the priors $\alpha$ and the posteriors $p_{bl}(y)$ of bid leakage. We discuss and verify applicability of \textit{parity} in Section \ref{sec-verify-assumption}.

\section{Empirical Results}\label{sec-results}

\subsection{Applying the strategy to the main sample.}\label{sec-results-strategy}

In this subsection we summarize our estimates of the prior and the posterior probabilities of bid leakage in the main sample of 3 and more participants.

\begin{table}[h]
  \centering
  \caption{Estimated proportion of corrupted auctions}
  \label{table-priors}
  \small
  \begin{tabular}{|c|c|c|c|}
  \hline
  \multicolumn{2}{|c|}{real}                                      & \multicolumn{2}{c|}{placebo}                                    \\ \hline
  NN                             & XGB                            & NN                             & XGB                            \\ \hline
  $0.089 \pm 0.011$ & $0.084 \pm 0.009$ & $0.002 \pm 0.001$ & $0.002 \pm 0.002$ \\ \hline
  \end{tabular}
\smallskip
\footnotesize
\begin{flushleft}
\textit{Notes}: In the column `real`, the original set Real(1,2) is treated as real, and the placebo set with dropped 1st places Placebo(2,3) is treated as placebo. In the column `placebo`, the placebo set with dropped 1st places Placebo(2, 3) is treated as real, and the placebo set Placebo(3,4) with dropped 1st and 2nd places is treated as placebo. The subcolumns `NN` and `XGB` denote different classifiers. The statistics are gathered by independently applying the strategy 5 times.
\end{flushleft}
\normalsize
\end{table}

The overall proportion of auctions susceptible to bid leakage in the sample is estimated as 8-9\% (Table \ref{table-priors}). This estimate is a little smaller than 10-11\% found previously in \citep{korovkin2018detecting}, which is likely to be due to our \textit{parity} assumption being more conservative than \textit{independence}. These results are discussed in more details in Section \ref{sec-applying-placebo}.

We also wish to describe the decision making mechanism learned by the two-stage strategy and the patterns it identifies as suspicious. Unfortunately, interpreting neural networks or ensembles of decision trees is a non-trivial task, whereas simpler classifiers lack expressive power. Incorporating the post-processing stage that involves density estimation further complicates the task. 

\begin{figure}[h]
    \centering
    \caption{Decision tree regression trained on bid leakage predictions.}
    \includegraphics[width=\linewidth]{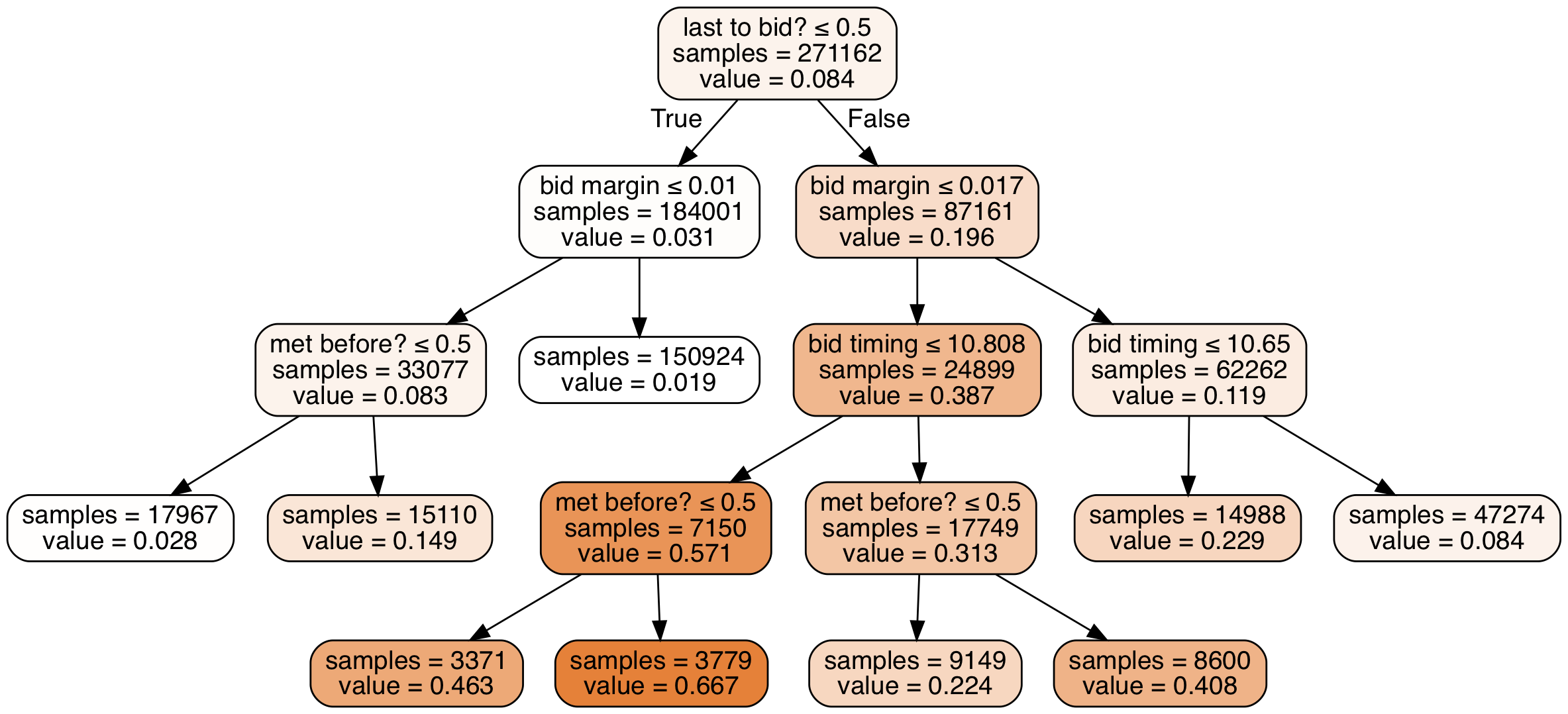}
    \label{fig:tree}
\footnotesize
\begin{flushleft}
\textit{Notes}: The tree is trained on the features from Table \ref{table-features} to predict the output of the bid leakage estimation strategy. In each node, the first line is the condition that optimally splits the data (absent in leaves), the second line is subsample size, and the third line is average bid leakage prediction in the subsample. Left arrows point to the subsamples that meet the conditions, and vice versa.
\end{flushleft}
\normalsize
\end{figure}

Instead of pursuing inherent interpretability, we propose a simple alternative. Specifically, we train a truncated decision tree (Fig. \ref{fig:tree}) to approximate the predictions of the strategy. While this approach does not uncover all the complex feature interactions, it provides insights into the major patterns that the strategy deems suspicious. Remarkably, these patterns coincide with the rational behaviour of the participant that received leaked bids (Section \ref{sec-introduction}) and the anomalies that we identified during data analysis (Section \ref{data-analysis}). The most suspicious auctions can be described by the following set of features (in order of decreasing importance): the winner bids last in the auction, within 1-2\% of the second bid, within the last 10 minutes of the auction, and has participated in the auction with the same procurer before. More details can be found in Figure \ref{fig:tree}.

To reiterate, this decision tree presents only the rule of thumb interpretation. The features on their own might be not informative about bid leakage; what matters is how more specific they are for the winners than for the runners-up. For example, small bid margins often come not from idiosyncratic information but from common shocks \citep{krasnokutskaya2011bid} and thus all bidders in an auction will have small bid margins. Only if the bid margin of the winners is on average smaller than the bid margin of the runners-up, will the classifier distinguish them. Similarly, as our auctions are frequent and time span is 4 years, both the winner and the runner-up are very likely to have met the procurer before. Only if the winners met the procurers earlier than the runners-up (in our sample), will the classifier distinguish them.

\subsection{Extending the strategy to the auctions with 2 participants.}\label{sec-results-2participants}

During the preprocessing stage, most of the excluded data are the auctions with 1 and 2 participants. The auctions with 1 participants cannot be analyzed with our strategy due to absence of the runners-up. However, the problem with the auctions with 2 participants is less severe. While both winners and runners-up are present in these auctions, a particular feature, namely \textit{bid margin}, cannot be computed for the runners-up due to absence of third places.\footnote{Estimating \textit{bid margin} for the runners-up in the auctions with 2 participants as the normalized difference with reserved price leads to target leakage, i.e. participants with zero margin are definitely runners-up.} One approach to extend our strategy to these auctions is to remove \textit{bid margin} from the feature set, but this comes at a loss of precision. We employ a different approach.

The procedure is the following. After bid leakage is estimated in the main sample, for each auction with two participants we find an auction in the main sample where the winner is most similar. We then assign the probabilities of bid leakage for the two winners as equal. Technically, this procedure is implemented as k-neighbours regression with $k = 1$,\footnote{The results are robust to the number of neighbours.} where the main sample and its bid leakage predictions are used as training data and target. To find the nearest neighbour, the same features as in Table \ref{table-features} (excluding \textit{number of participants}) are used. Since Euclidean distance is sensitive to scale, the features are preemptively normalized. The results are reported in Table \ref{table-priors-2part}.

\begin{table}[h]
  \centering
  \caption{Estimated proportion of corrupted auctions with 2 participants}
  \label{table-priors-2part}
  \small
  \begin{tabular}{|c|c|c|c|}
  \hline
  \multicolumn{2}{|c|}{real}                                      & \multicolumn{2}{c|}{placebo}                                    \\ \hline
  NN                             & XGB                            & NN                             & XGB                            \\ \hline
  $0.109 \pm 0.006$ & $0.100 \pm 0.010$ & $0.002 \pm 0.002$ & $0.002 \pm 0.003$ \\ \hline
  \end{tabular}
\smallskip
\footnotesize
\begin{flushleft}
\textit{Notes}: The table is analogous to Table \ref{table-priors} but for the sample of auctions with 2 participants.
\end{flushleft}
\normalsize
\end{table}

It is difficult to judge how precise the outlined procedure is. Since feature distributions vary with the number of participants, the estimate for this sample might be biased. On the one hand, a fair winner is the more likely to bid last the less the participants, which might lead to overestimation. On the other hand, the fewer the participants the less likely the bids of fair winners and runners-up are to be close, which might lead to underestimation. It is unclear which of these factors is dominating, if any. Still, the proportion estimates (Table \ref{table-priors-2part}) seem plausible, especially considering the above-average predictions for the auctions with low number of participants in the main sample (Fig. \ref{figure-bl-by}).

\subsection{Connection of bid leakage predictions with auction characteristics}

\begin{figure}[h]
    \centering
    \caption{Bid leakage probability aggregated by auction characteristics}
    \label{figure-bl-by}
    \includegraphics[width=\textwidth]{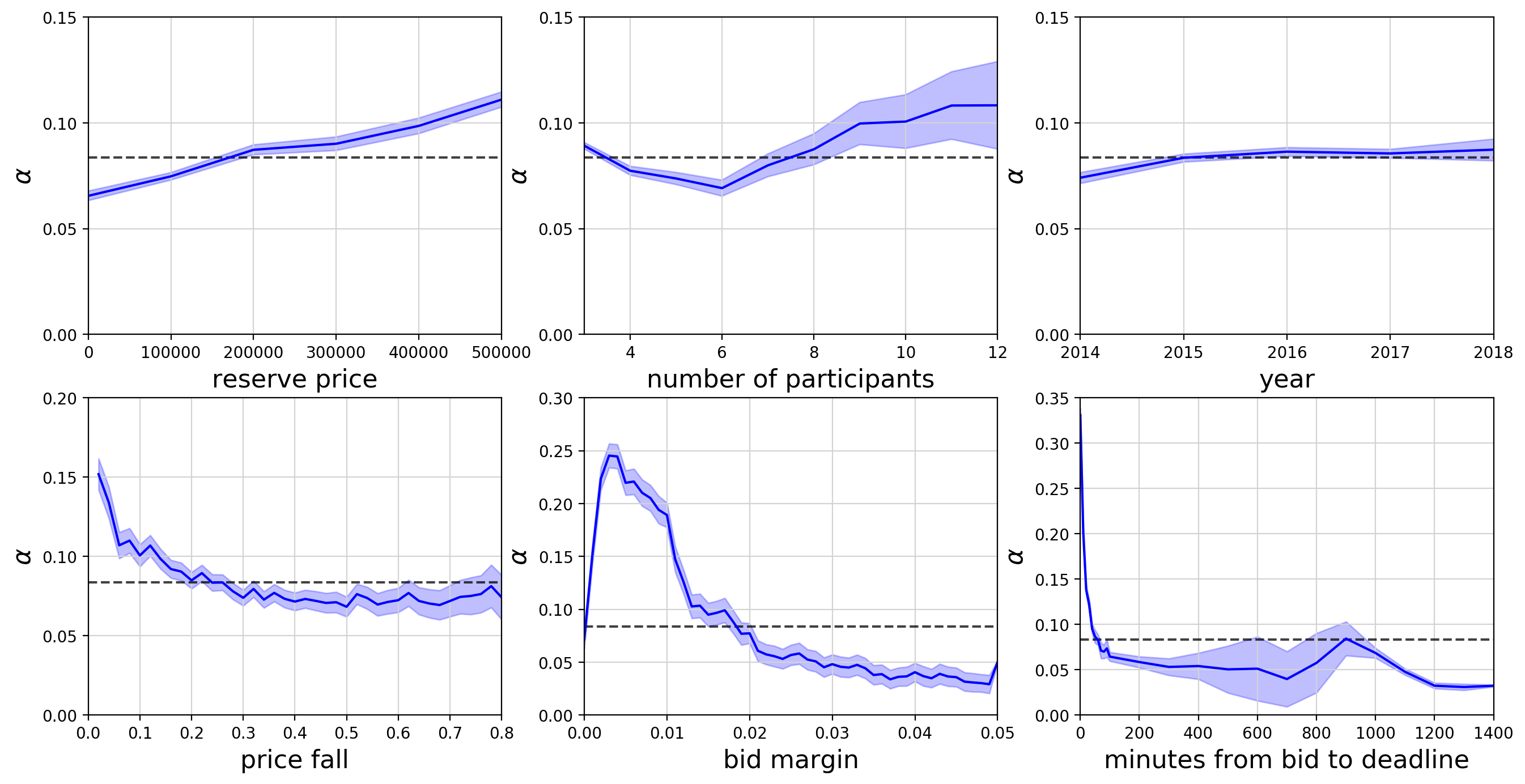}
\footnotesize
\begin{flushleft}
\textit{Notes}: Correlations with $\alpha$ for all plots are statistically significant on 0.0001 level. Correlations on the top row are positive and on the bottom row are negative. Filled regions represent 0.95 confidence interval. Dashed horizontal lines are plotted on the level of the prior probability. The last two plots correspond to the features `bid margin` and `bid timing` that are fed to the winner's classifier. As discussed in notes to Table \ref{table-features}, these features are clipped from above.
\end{flushleft}
\normalsize
\end{figure}

The probability of bid leakage for specific subsamples provides insights into the mechanisms behind bid leakage. We present these results in Fig.\ref{figure-bl-by}.

The top left diagram in Fig.\ref{figure-bl-by} demonstrates that as the reserve price increases, the bid leakage is slightly more likely to be observed. This is very natural and might have two explanations. First, a higher reserve price for a contract gives higher incentives to organize risky corruption schemes. Second, the better organized corruption schemes allow setting a higher reserve price in order to maximize the surplus.

The top center diagram in Fig.\ref{figure-bl-by} demonstrates that the extreme numbers of bidders on average corresponds to higher probability of bid leakage. To make bid leakage profitable it should be accompanied with a boosted reserve price. The higher reserve price attracts more bidders. If the auctioneer is able to deter entry (e.g., by requiring certification, by making the announcement difficult to find, by having bad reputation), only few bidders enter. If the auctioneer is not able to deter entry, many bidders enter.

The top right diagram illustrates that bid leakage is consistent throughout the years.

\begin{figure}[t]
\centering
    \caption{Estimated prevalence of bid leakage in regions of Russia}
    \label{figure-russia}
    \includegraphics[width=\textwidth]{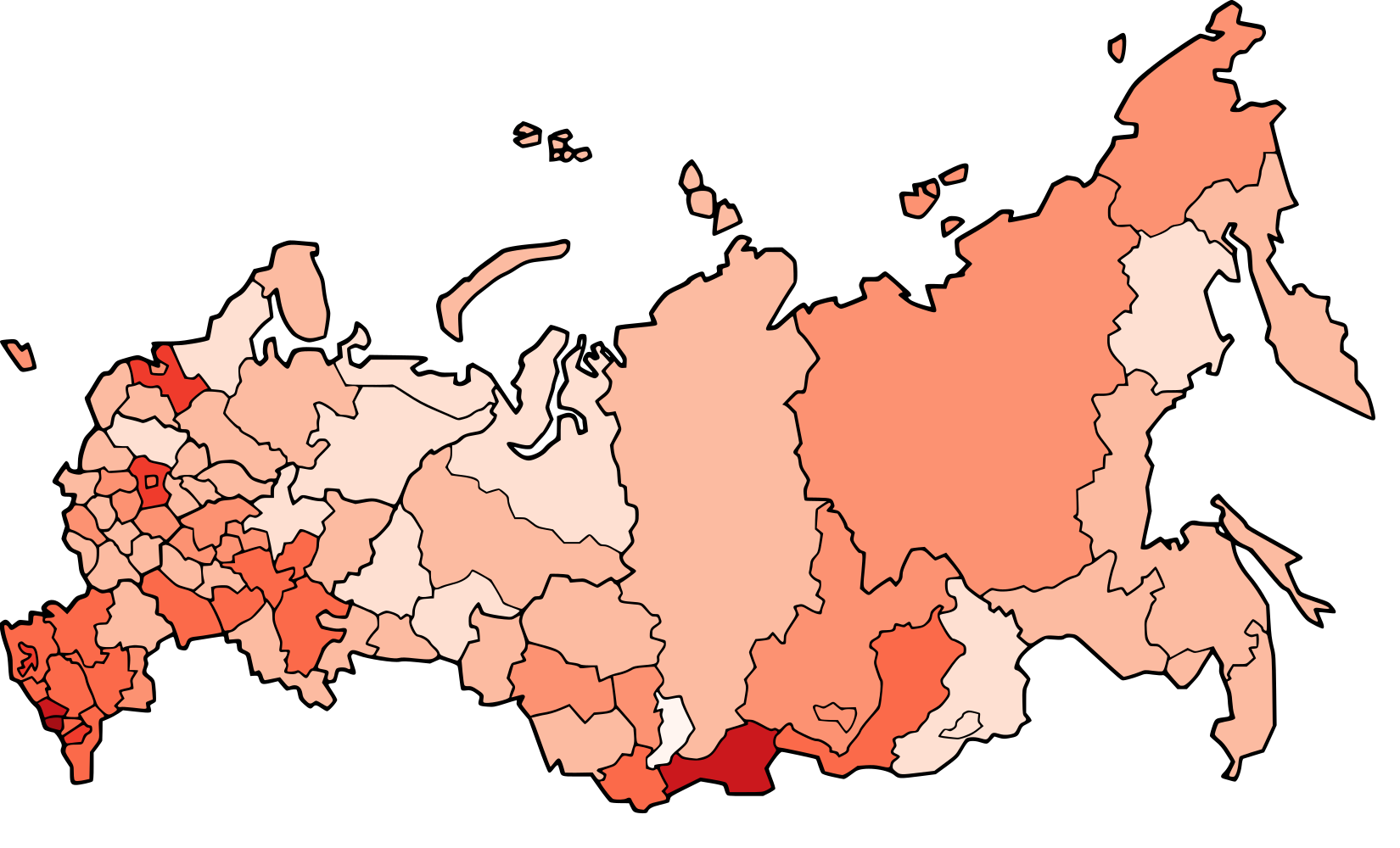}
    \includegraphics[scale=0.55]{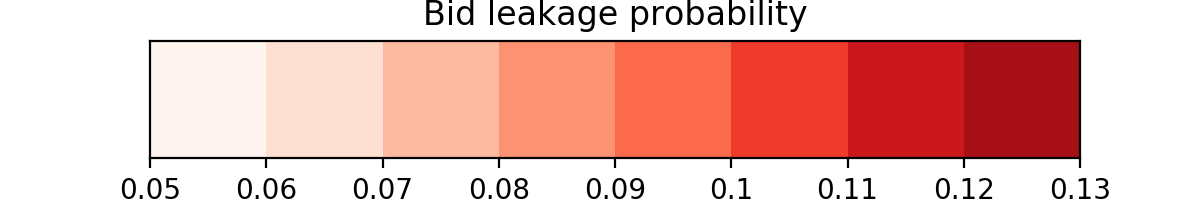}
\end{figure}

The bottom left diagram in Fig.\ref{figure-bl-by} shows a stark relation of bid leakage and the price fall -- the ratio of the final price and the reserve price. Higher probability of bid leakage is associated with higher final prices as bid leakage hampers competition.

The bottom center diagram in Fig.\ref{figure-bl-by} demonstrates that lower bid margin is associated with higher probability of bid leakage, except for the bid margins below 0.2\% of the reserve price. (This is around few hundreds of roubles on average. Lower bid margins might seem suspicious and thus are not chosen.)

The bottom right diagram in Fig.\ref{figure-bl-by} shows that the (estimated) favored participants are the more likely to submit their bids the closer is the deadline, which was anticipated. 

Finally, Fig.\ref{figure-russia} presents the cross-regional average probabilities of bid leakage. Perhaps surprisingly, the variance between different regions is rather large. However, these results are consistent with other measures of corruption available for Russian regions, such as electoral fraud (see, e.g., \cite{mebane2009comparative,bader2015explains}).

\subsection{Increase of procurement costs due to bid leakage.}\label{sec-results-welfare}

\begin{figure}
    \centering
    \caption{Price fall regressed on bid leakage probability}
    \label{fig:welfare_loss}
    \includegraphics[width=\textwidth]{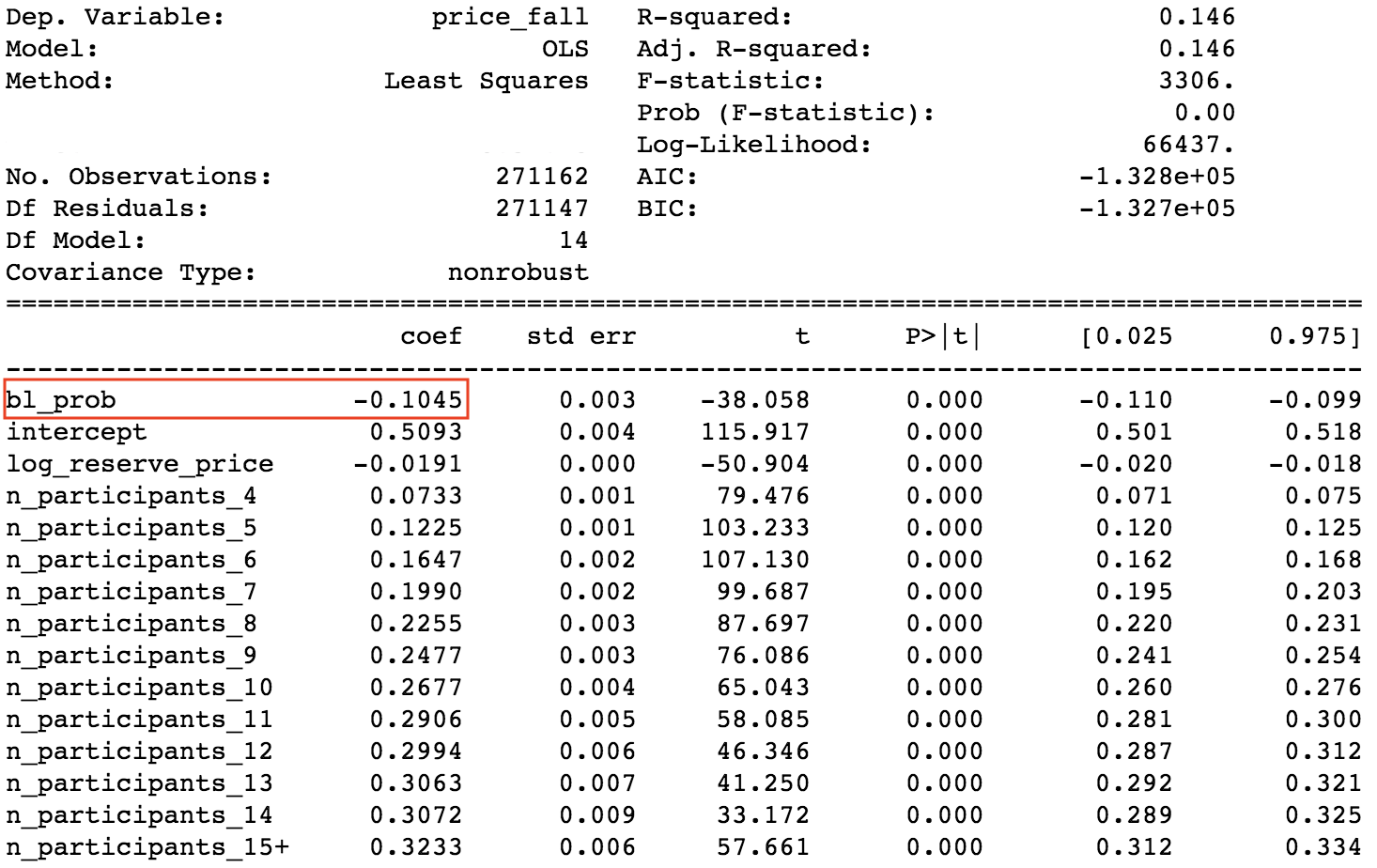}
\footnotesize
\begin{flushleft}
\textit{Notes}: The dependent variable is the normalized price fall defined as $\frac{r - b_1}{r}$, where $r$ is reserve price, $b_1$ is winner's bid. The regressor of interest is `bl\_prob`, i.e. the predicted probability of bid leakage. The estimated coefficient $-0.1045$ can be interpreted as the average decrease of price fall in a corrupted auction in comparison to a fair one. As control variables we use logarithm of reserve price and one-hot encoded number of participants. Including geographical region as a control variable insignificantly changes the `bl\_prob` coefficient to $-0.1025$. To test for multicollinearity, we calculate variance inflation factors (VIFs) for the regressors. The highest VIF among the regressors is estimated as $1.13$, which signals absence of multicollinearity by the rule of thumb $VIF < 5$. The experiments were conducted using the statsmodels package \citep{statsmodels}.
\end{flushleft}
\normalsize
\end{figure}

To estimate changes of procurement costs caused by bid leakage, we first construct ordinary least squares regression of the price fall on the predicted probability of bid leakage. For results and technical details, refer to Figure \ref{fig:welfare_loss}.

The fitted regression is then applied to the initial dataset it was trained on, and to the modified fair dataset where the probability of bid leakage is manually set to zero for all auctions. This allows us to estimate the difference between the expected total prices of auctions depending on whether bid leakage is present or absent. The difference is estimated at 520 millions roubles when normalized price fall (in percentage of reserve price) is used as dependent variable, and at 470 millions roubles when price fall is not normalized (in roubles). These estimates correspond to around 1\% of the sum of reserve prices and around 1.5\% of the sum of final prices.

\section{Validation}\label{sec-validation}

It is inherently difficult to evaluate performance of a method with no labeled data at hand. A desirable alternative to proper a validation could be to estimate precision of the method on a small sample of auctions proven to be corrupted, but such sample is unavailable to us. Fortunately, much of the job can be done using the placebo datasets. First, placebo allows us to explicitly check \textit{independence} and \textit{parity} assumptions. We find that \textit{independence} does not hold, whereas \textit{parity} is plausible. Second, we measure overestimation of our strategy by applying it to the fair placebo auctions. As expected, close to no corruption is found there. 

We use two placebo datasets in our analysis, one with dropped winners and another with dropped winners and runners-up, denoted as Placebo(2,3) and Placebo(3,4) respectively. Similarly, Real(1,2) denotes the original dataset with no dropped participants. The relationship between the two placebo datasets is supposed to mimic such relationship between Real(1,2) and Placebo(2,3) in the absence of corruption. In other words, similarity of the placebo datasets implies that the difference with the original dataset is explained by bid leakage.


\subsection{Verifying Parity and Independence Assumptions}\label{sec-verify-assumption}


In order to provide some intuition behind the \textit{parity} assumption, we begin by deconstructing it. First of all, note that \textit{parity} is a statement about `distributions of the winner's classifier' predictions $y(x)$, rather than `distributions of the initial features` $x$. This is a subtle but important distinction, since the latter formulation is stronger. Still, if the assumption was formulated about $x$ rather than $y(x)$, it could roughly be translated as `the difference between the pdfs of first and second order statistics of feature distributions is equal to this difference between second and third order statistics`. It might be unclear why such assumption would hold for complex feature distributions if it does not hold for such simple distributions as uniform or normal. This is the point where the distinction between $x$ and $y(x)$ becomes relevant.  

On the one hand, if \textit{parity} does not hold for $x$, it may still hold for $y(x)$ due to imperfectness of the classifier. Even if $\Delta_{12}(x)$ differs from $\Delta_{23}(x)$, the classifier might not accurately reflect this difference. As an extreme example, a random-guess classifier (predicts constant $y$ independent of $x$) would yield such predictions that $\Delta_{12}(y)=\Delta_{23}(y)=0$ for any feature distributions.

On the other hand, and more importantly, both \textit{independence} and \textit{parity} can be statistically validated when formulated for $y(x)$. It is difficult to validate whether $\Delta_{12}(x)=\Delta_{23}(x)$ for two major reasons. The first reason is the curse of dimensionality that hinders density estimation of multi-dimensional feature distributions of $x$. In contrast, the curse of dimensionality is not an issue for estimation of one-dimensional pdfs and cdfs of $y(x)$. The second reason is the lack of access to $\Delta_{12}(x)$ or $\Delta_{12}(y)$ due to contamination of the real data with corrupted auctions. In order to verify \textit{parity}, we instead statistically validate whether $\Delta_{23}(y)=\Delta_{34}(y)$, which in turn would signal equality of $\Delta_{12}(y)$ and $\Delta_{23}(y)$. Similarly, we verify \textit{independence} by testing whether $\Delta_{23}(y)=0$ rather than $\Delta_{12}(y)=0$. 

To test whether $\Delta_{23}(y)=0$, i.e. whether the classifier's predictions for winners and runners-up of placebo auctions are generated from the same distribution, we perform two-sample Kolmogorov-Smirnov (KS) test \citep{lilliefors1967kolmogorov}. The statistic in KS test is the supremum of the absolute difference between the empirical cdfs of two samples ($\sup \left| F_{1_{(2,3)}}(y) - F_{2_{(2,3)}}(y) \right|$, where $F$ denotes cdf). The resulting p-value is equal to $10^{-28}$, which rejects the null hypothesis of equality of $f_{1_{(2,3)}}(y)$ and $f_{2_{(2,3)}}(y)$. Regarding \textit{parity}, testing equality of $\Delta_{12}(y)$ and $\Delta_{23}(y)$ is not as straightforward. To the best of our knowledge, there is no statistical test that validates equality of differences between distributions. As an approximation, we perform KS test with the statistic chosen as the supremum of the absolute difference of the differences between the empirical cdfs ($\sup \mid (F_{1_{(2,3)}}(y) - F_{2_{(2,3)}}(y)) - (F_{1_{(3,4)}}(y) - F_{2_{(3,4)}}(y)) \mid$). The resulting p-value is equal to $0.006$, which is significant on $0.01$ but insignificant on $0.001$ probability threshold. According to this result, we cannot confidently conclude that \textit{parity} holds. However, \textit{parity} is tens of magnitudes more likely to hold than \textit{independence}. 

\begin{table}[t]
\centering
\small
\caption{Classifier's performance on the original and the placebo datasets}
\label{table-metrics}
\begin{tabular}{|c|c|c|c|}
\hline
         & Real(1,2) & Placebo(2,3)      & Placebo(3,4)      \\ \hline
Accuracy & $0.5383 \pm 0.0008$        & $0.5090 \pm 0.0002$ & $0.5061 \pm 0.0005$ \\ \hline
ROC-AUC  & $0.5520 \pm 0.0007$        & $0.5103 \pm 0.0002$ & $0.5058 \pm 0.0005$ \\ \hline
\end{tabular}
\smallskip
\footnotesize
\begin{flushleft}
\textit{Notes}: mean and standard deviation statistics of scores are obtained by training 5 randomly initialized classifiers. For all columns, the classifiers are trained on Real(1,2). Given the potentially low corruption rate, the classifier's low performance is not surprising, provided that the performance on the placebo is lower then on the original data. The scores are reported for the classifier based on neural network; the classifier based on the gradient boosting performs similarly.
\end{flushleft}
\normalsize
\end{table}

\begin{figure}[t]
    \centering
    \caption{Differences of pdfs of classifier's predictions}
    \includegraphics[width=\linewidth]{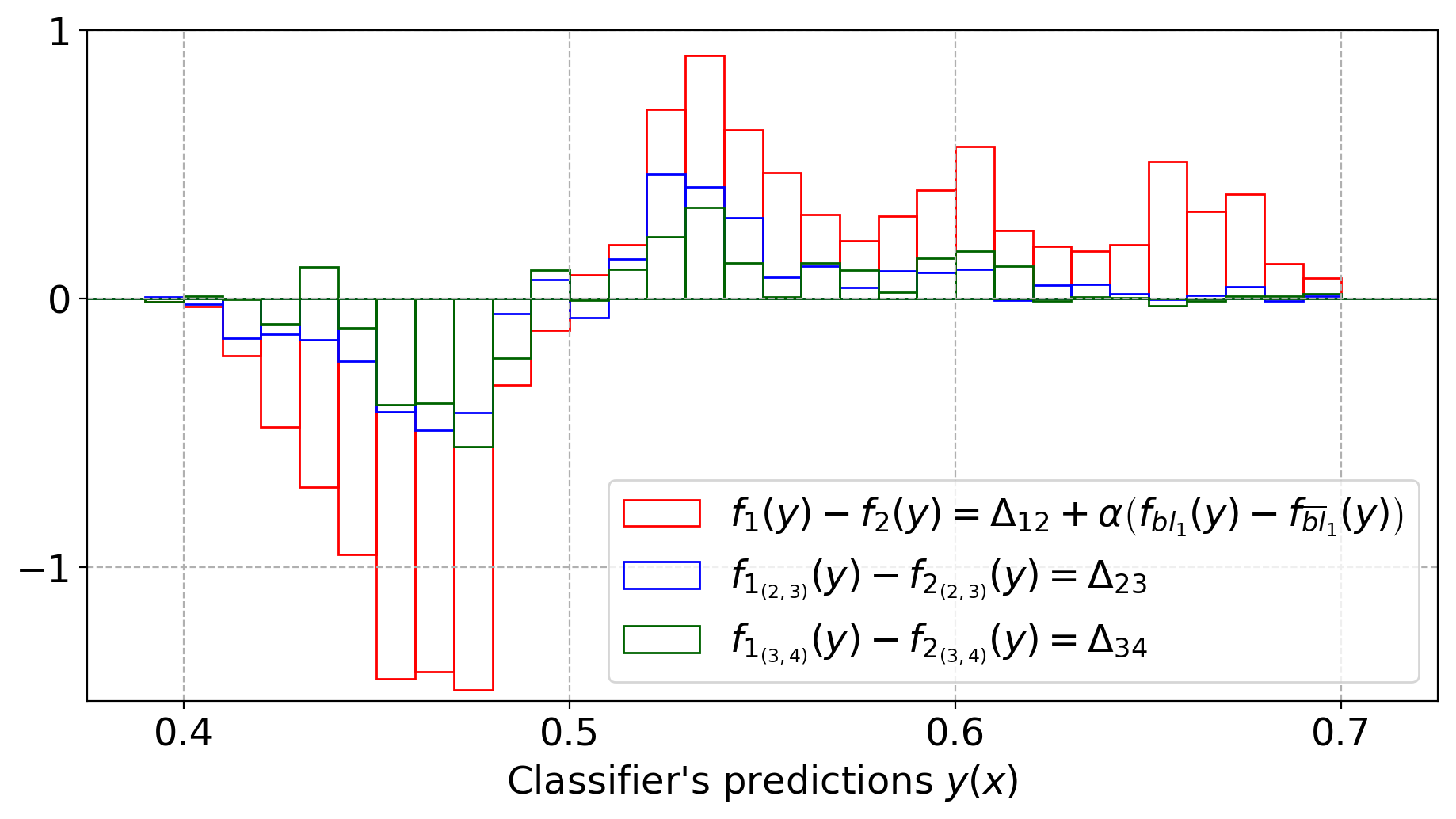}
    \label{figure-delta}
\footnotesize
\begin{flushleft}
\textit{Notes}: Each pdf is estimated as a normalized histogram. Red bins denote pdf difference for original data, whereas blue and green bins denote pdf differences for placebo data with dropped winners and with dropped winners and runners-up, respectively. The width of each bin is equal to 0.01.
\end{flushleft}
\normalsize
\end{figure}

Additionally, we provide indirect evidence of \textit{parity} being more plausible than \textit{independence} by comparing the assumptions in terms of absolute values. To this end, we report the classifier's performance (Table \ref{table-metrics}), as well as the differences of pdfs of the classifier's predictions for the winners and the runners-up, explicitly estimated as differences of normalized histograms (Figure \ref{figure-delta}).

We first discuss the classifier's performance (Table \ref{table-metrics}). If \textit{independence} holds, we expect the classifier to perform on placebo data as well as random guess, i.e. the classifier should not distinguish between fair winners and runners-up based on time-related features (while we also use a bid-related feature `bid margin`, the observations below persist if this feature is excluded). Instead, we observe that the classifier has above-random performance on both Placebo(2, 3) and Placebo(3,4) datasets. On the other hand, if \textit{parity} holds, we only expect the classifier to perform identically on the two placebo datasets. While this is also not the case, the classifier's performance on the placebo datasets is more similar than the classifier's performance on any of the placebo datasets compared with a random-guess classifier. In other words, even if \textit{parity} is violated, \textit{independence} is violated more. At the same time, the classifier's performance on the real dataset is considerably higher than on the placebo datasets, which is expected in the presence of bid leakage.

Similar conclusions can be drawn by comparing differences of pdfs (Figure \ref{figure-delta}). If \textit{independence} holds, we expect $\Delta_{23}(y) = \Delta_{34}(y) = 0$ for all $y(x)$, whereas \textit{parity} only implies $\Delta_{23}(y) = \Delta_{34}(y)$. Observe that these pdf differences for placebo data sets are clearly non-zero on the [0.4, 0.7] probability interval, but are similar, albeit not identical. At the same time, the pdf difference for the real auctions is considerably more vivid due to the presence of bid leakage.

Finally, we attempt to verify the \textit{independence} assumption with the same method as \cite{korovkin2018detecting}. To this end \cite{korovkin2018detecting} exclude all winning last bids and then test if being last predicts a lower bid. They find the opposite effect (in this subsample earlier bids are lower) of a much lower size. We replicate this observation with our dataset.

However, if we exclude \textit{all} winning bids, we find that later bids are more likely to be smaller. That is, in the imaginary auction with only 2nd and 3rd lowest bids, the lower 2nd bid is on average submitted later. This is also true when we exclude winners and 2nd lowest bids and compare 3rd and 4th latest bids: the lower 3rd bid is on average submitted later. This effect is also evident if the bid timings of winners and runners-up are compared with the population (Table \ref{table-data}).  

\subsection{Applying the Method to Placebo}\label{sec-applying-placebo}

To check our method for overestimation, we apply it to the placebo dataset, where no corruption is ideally to be identified (Table \ref{table-priors}). Specifically, we run the procedure outlined in Algorithm \ref{algorithm-strategy}, with Placebo(2,3) provided as the possibly-corrupted dataset, and Placebo(3,4) provided as the fair placebo. This yields a negligible estimate of percentage fractions, almost two magnitudes less than in the original dataset. Moreover, no auctions in Placebo(2,3) are actually labeled as corrupted, as all the predicted probabilities are below the 0.5 threshold. To compare, around 1.5\% of the auctions in the original dataset are labeled as corrupted.

\section{Conclusions}\label{sec-conclusions}

We study first-price sealed-bid auctions and identify the auctions corrupted with bid leakage. The first stage of our strategy is to build a classifier that successfully distinguishes the winners from the runners-up in the corrupted auctions but not as well in the fair auctions. In the second stage we process the classifier's predicted probabilities of winning for the winner and for the runner-up of each auction into the probability that the auction is corrupted. We apply our  estimation strategy to the Russian procurement data between January 2014 and March 2018 containing around 600000 auctions after the preprocessing.

We estimate the share of the corrupted auctions in the dataset as around 9\%. We believe that this estimate is conservative due to the assumptions we make. First, we are only concerned with effective bid leakage, that is if the bids are leaked to the favored participant, she inevitably wins. Consequently, when the classifier selects a runner-up, we assume it to be a mistake. Second, instead of \textit{independence} we use \textit{parity}, an assumption that is much more plausible (Section \ref{sec-verify-assumption}) and that lowers the proportion estimate by around a third.

The assumption that only the winners are corrupted is incidentally verified together with \textit{parity}. If the losing bids are corrupted, then it is natural to expect most of them to be runners-up (and not the other losing participants). For instance, one fair bid being submitted late enough in order not to be leaked is more probable than a scenario with two or more such bids. In this case, we would expect the pdf ratios $\Delta_{23}$ and $\Delta_{34}$ to differ substantially. However, this is not the case (Figure \ref{figure-delta}).

We must state our reservations regarding the alternative interpretations of our results. Under some assumptions, the patterns studied here can also come from other forms of illegal behavior, such as collusion. Imagine that all bidders collude and want to commit to it. To this end, the supposed winner can bid last so that the others cannot deviate. If the cartel includes all participants in the market, this behaviour may induce the same pattern as in bid leakage: bidding last is associated with winning. However, there are also two differences: as there is no need to bid close to the deadline, the feature ``bid timing'' would not be informative, same as the feature 
``met before?'', which represents the relationship with the auctioneer but not necessarily the relationship with the competitors. Disentangling this and the other types of behavior is a subject of further research. 

We estimate the direct welfare loss as 1.5\% of the total sum of contracts distributed via the auctions in our sample. This number might not strike as too high, especially given that the auctions are small. But these direct losses might cause further corruption. The quality of the goods and services procured through corrupt auctions might be lower. Future auctions and other procurement procedures might also be corrupted using these or other schemes. One malevolent experience might be contagious and spread to other procurers and bidders like a virus.

\bibliographystyle{informs2014}
\bibliography{bid-leakage-bibligrophay}


\end{document}